\documentstyle[prl,aps,multicol,epsf]{revtex}
\begin{document}
\draft
\title{Dynamics of an SO(5)-symmetric ladder model}
\author{R. Eder$^1$, A. Dorneich$^1$, M. G. Zacher$^{1,2}$, W. Hanke$^1$, 
Shou-Cheng Zhang$^3$}
\address{$^1$Institut f\"ur Theoretische Physik, Universit\"at W\"urzburg,
Am Hubland,  97074 W\"urzburg, Germany\\
$^2$ Institut for Solid State Physics, University of Tokyo,
Tokyo, Japan\\
$^3$ Department of Physics, Stanford University, Stanford, USA
}
\date{\today}
\maketitle
\begin{abstract}
We discuss properties of a recently proposed exactly $SO(5)$-symmetric
ladder model. In the strong coupling limit we demonstrate how the
$SO(3)$-symmetric description of spin ladders in terms of bond Bosons
can be upgraded to an $SO(5)$-symmetric bond-Boson model, which provides
a particularly simple example for the concept of $SO(5)$ symmetry.
Based on this representation we show that antiferromagnetism on one hand 
and superconductivity on the other hand can be understood as
condensation of either magnetic or charged Bosons into an RVB vacuum.
We identify exact eigenstates of a finite cluster with general 
multiplets of the $SO(5)$ group, and
present numerical results for the single particle, spin and charge
spectra of the $SO(5)$-symmetric model and identify `fingerprints' of
$SO(5)$ symmetry in these. In particluar we show that $SO(5)$
symmetry implies a `generalized rigid band behavior' of the
photoemission spectrum, i.e. spectra for the doped case are 
rigorously identical to spectra for spin-polarized states at half-filling.
\end{abstract}
\begin{multicols}{2}

%%%%%%%%%%%%%%%%%%%%%%%%%%%%%%%%%%%%%%%%%%%%%%%%%%%%%%%%%%%%%%
\section{Introduction}
%%%%%%%%%%%%%%%%%%%%%%%%%%%%%%%%%%%%%%%%%%%%%%%%%%%%%%%%%%%%%%

It has recently been proposed\cite{science}
that the antiferromagnetic (AF) and
superconducting (SC) phases of the high-$T_{c}$ cuprates are unified by an 
$SO(5)$ symmetry principle. 
A remarkable degree of continuity between insulating and doped phase is
indeed supported by a considerable body of numerical 
evidence\cite{EderOhtaShimozato,charge,Preuss,Nishimoto}.
Further support for this proposal came from
numerical investigations, which demonstrated that the low-energy excitations
of physical models for the high-$T_{c}$ materials, i.e. $t-J$ and
Hubbard models, can be classified in terms of an $SO(5)-$symmetry multiplet
structure\cite{Meixner,ede97}. 
In these two-dimensional ($2D$) microscopic models, the
undoped situation - in agreement with the experimental situation in the
cuprates - corresponds to that of a Mott insulator with broken $SO(3)$ or
spin rotational symmetry: long-range AF order is realized. The $SO(5)-$%
symmetry principle then tells us how this long-range magnetic order and the
accompanying low-energy spin excitations are
mapped onto the corresponding off-diagonal long-range SC order and the
low-energy ``Goldstone bosons" (the $\pi -$mode) in the doped 
situation\cite{Demler,Meixner,ede97}.\\
However, there exists also a second class of Mott-type insulators without
long-range AF order, i.e. spin liquids, which have a gap to spin excitations.
There is growing experimental evidence that they are also intimately
related to the physics of high-$T_{c}$ compounds: not only do these
compounds show above the Ne\`{e}l temperature and superconducting transition
temperature at small dopings signs of such a spin gap, but there exist also
copper-oxides with a $Cu0_{2}$ plane containing line-defects, which result
in ladder-like arrangements of $Cu-$atoms\cite{Hiroi,Azuma,Uehara,DagottoRice}.
These systems can be
described in terms of coupled two-leg ladders\cite{DagottoRice}, 
which exhibit a spin gap
in the insulating compound and thus belong to the spin-liquid Mott-insulator
variety. Also the related ``stripe phases" of the $2D$ $Cu0_{2}$ planes in
the cuprate superconductors have recently received considerable 
attention\cite{Tranquada,Zaanen}. 
In these systems, the apparent connection between the spin gap and
superconductivity must be explained.\\ 
In order to illustrate how the $SO(5)$ theory can, in principle, cope with
this challenge, an exactly $SO(5)$ invariant ladder model has recently been
constructed\cite{S-Z-H}. It was shown that a 2-leg ladder with entirely local
interactions, i.e. an on-site interaction $|U|>>t$, where $t$ denotes the
chain hopping, an intra-rung interaction $|V|>>t$ and a magnetic
rung-exchange interaction $J$ can have $SO(5)$ symmetry if these
interactions are related to each other in a specific way, i.e. $J=4(U+V)$.
The ground-state phase diagram of this model was determined in strong
coupling and, in particular, a regime identified, where the strong-coupling
ground state is a spin-gap insulator. In addition, the spin-gap magnon mode
of the Mott insulator was shown to evolve continuously into the $\pi -$%
resonance mode of the superconductor.\\ 
However, two key questions remained open, the first being the relationship
of this $SO(5)$ ladder to the ``physical" $t-J$ or Hubbard ladders and the
second regarding the connection to the other variety of Mott insulators,
i.e. the ones with long-range AF order.\\
With regard to the first question, progress was recently made in the regime
of weak-coupling: using the weak-coupling renormalization
group method, two independent 
works\cite{Enrico,Fisher}
have recently demonstrated that rather generic ladder models at
half-filling flows to an $SO(5)$ symmetric fixed point.\\
In the present work, we try to attack both questions in the
experimentally relevant intermediate and strong-coupling regimes. The basic
idea is to derive the $SO(5)$ concept for a ladder or, more generally, for a
spin liquid with an ``RVB vacuum" instead of an ``Ne\`{e}l vacuum"
involving a physically appealing new construction: an effective $SO(5)$
invariant low-energy Hamiltonian is constructed in terms of a bond operator
representation. In strong coupling, this is a well-established concept for
spin ladders\cite{Gopalan,ladder}
to account quantitatively for the spin excitations of Heisenberg ladders (up
to the physically most relevant case of isotropic couplings in the rungs and
along the ladder). The spin excitations are described in terms of triplet
fluctuations around the RVB vacuum. The new idea is to combine this $SO(3)-$%
symmetric description of the spin degrees of freedom , i.e. the triplet
excitations, with the charge (hole) dynamics. This is accomplished by
introducing two new Bosons on a rung which stand for charge fluctuations,
i.e. empty and fourfold occupied rungs. When combined into a 5-dimensional
Boson-vector, the dynamics of spin and charge excitations is captured by a
manifestly $SO(5)$ invariant Hamiltonian.\\ 
It is shown that this formulation in terms of triplet and hole fluctuations
around an ``RVB vacuum" allows for a physically transparent demonstration
of the corner stone in $SO(5)$ theory, i.e. that AF and SC are ``two faces
of one and the same coin". By starting from this ``RVB vacuum", which
represents the spin liquid state at half-filling, we demonstrate that an AF
ordered state can be generated by forming a coherent state 
\[
|\psi >\text{ }\sim e^{\lambda t_{z}^{+}(q=\pi )}|\Omega>, 
\]
which corresponds to $z-$like triplets condensed into the $q=\pi $ state.
However, in the $SO(5)$ theory, the $z-$like triplet with momentum $\pi $
and the hole pair with momentum $0$ are components of one and the same 
$SO(5) $ vector. They are rotated into each other by the $SO(5)$ generating
operator $\pi $. This implies that the above coherent state with condensed
triplets can be $SO(5)-$rotated into a corresponding coherent state with $%
t_{z}^{+}(q=\pi )$ replaced by the (hole-) pair creation operator $\Delta
^{+}(q=0).$ This state corresponds to hole pairs condensed into the $q=0$
state, i.e. a superconducting state. In other words: both the AF and the SC
state can be viewed as a kind of condensation out of the RVB state, or the
spin liquid. If the so constructed AF state is the actual ground state at
half-filling, then this physically very appealing $SO(5)$ construction
yields automatically the ground state in the doped situation, i.e. the SC
state.\\
This construction can also shed light on the second of the above questions,
namely the interrelation between the spin gap Mott-insulator and
superconductivity. The construction presented here rests on the special
geometry of the 2-leg ladder, which suggests a unique ``RVB vacuum", from
which the AF coherent state can be obtained. In $2-D$, such a unique vacuum
does, in general, not exist, unless one takes the growing experimental
evidence for some form of spatial inhomogenities such as 
stripes\cite{Tranquada,Zaanen} into
account. However, even for a translationally invariant state, up to an
additional statistical average over all possible dimer (singlet) coverings
of the plane, the analogue of the triplet-like excitation in the ladder can
still be generated: here singlet-dimers are substituted by triplet-dimers and
this excited dimer propagates through the $2D-$system. Again, the $SO(5)$ $%
\pi -$operator converts excited dimers with momentum $(\pi ,\pi )$ into $%
d_{x^{2}-y^{2}}-$symmetry hole pairs with momentum $(0,0).$ A description of
spin excitations in $2D$ in terms of Boson-like excited dimers, as for the
ladder considered here, may thus be a natural starting point for clarifying
the role of the spin gap (for which the ladder is a ``toy model") and thus
the role of the spin-liquid Mott-insulators for superconductivity, in
general. This will be discussed in a forthcoming 
paper\cite{tobepub}.\\ 
The first question, i.e. the adiabatic connection between the $SO(5)$
symmetric model and ``physical" ladder models, such as the $t-J$ model, is
answered numerically in the present work. This is achieved by a kind of
``Landau mapping" of the single-and two-particle excitations of the $SO(5)$
ladder model onto the corresponding excitations of the $t-J$ ladder.\\
As mentioned above, the empty-fourfold rung fluctuation involves, in
principle, a high energy (of order $\simeq 2U$). In the $SO(5)$ description,
the Hamiltonian is supplemented by a term which contains density and
exchange interactions, and which physically pulls down the empty-fourfold
fluctuation to be degenerate with the triplet-triplet fluctuation. It has
been shown in Ref. \cite{S-Z-H} that this can be achieved already for 
purely local
interactions $V$ and $J$ within the rungs by choosing $4(U+V)=J$. We
demonstrate numerically that, and give physical reasons why, this at first
sight rather unphysical parameter condition
nevertheless gives in fact an astonishingly good mapping of
the $SO(5)$ model onto the $t-J$ model for the low-energy $\omega $ versus
momentum $q$ dynamics. These findings strongly support the physical
relevance of the $SO(5)$ description in physical ladder models, such as
Hubbard and $t-J$ models, which are of the spin-liquid Mott insulator
variety.\\ 
The paper is organized as follows: Sec. II describes the construction of an 
$SO(5)$ symmetric ladder model in terms of bond-triplet and -charge bosons,
which are unified into a common $SO(5)$ vector $\overrightarrow{(t)}_{i}$.
It also contains the coherent-state description, where an AF ordered state
with staggered magnetization in $z-$direction is generated by a condensation
of $z-$like triplet excitations. When the triplets are rotated (via the $\pi
-$operator) onto the corresponding hole pairs (with $d_{x^{2}-y^{2}}$
symmetry), a $d-SC$ state results. Sec. III then contains a general
representation of the $SO(5)$ multiplet structure in terms of irreducible
representations in the $Q-S_z$ plane (here
$Q$$=$$(N_e-N)/2$ with $N_e$ the
number of electrons and $N$ the sunmber of sites\cite{science}
is the pair density and $S_{z}$ the $z$-component of total
spin). We consider here the previously studied multiplets of
diamond shape\cite{ede97} 
(corresponding to even numbers of electrons or holes, i.e.
electron or hole pairs) as well as multiplets evolving to a square shape. The
latter correspond to odd numbers of charge carriers and are thus required
for building up selction rules for photoemission. Sec. IV verifies these
selelction rules in exact diagonalizations of $2\times 6$ $SO(5)$ ladders.
Similarly, sec. V numerically illustrates the spin and charge dynamics of
the exact $SO(5)$ model and checks corresponding selection rules for spin-
and charge correlation functions. Sec. VI discusses the ``Landau mapping"
and adiabatic connection to the physical $t-J$ model, and sec. VII gives a
short conclusion.
%%%%%%%%%%%%%%%%%%%%%%%%%%%%%%%%%%%%%%%%%%%%%%%%%%%%%%%%%%%%%%
\section{Elementary excitations of the ladder}
%%%%%%%%%%%%%%%%%%%%%%%%%%%%%%%%%%%%%%%%%%%%%%%%%%%%%%%%%%%%%%
We begin with the standard Hubbard model on a 2-leg ladder, i.e.
\begin{equation}
H = -\sum_{i,j} t_{ij}
c_{i,\sigma}^\dagger c_{j,\sigma}^{}
+ U \sum_i (n_{i,\uparrow}-\frac{1}{2})(n_{i,\downarrow}-\frac{1}{2}).
\end{equation}
The nearest-neighbor hopping integral along the legs will be called 
$t$, the one within the rungs $t_\perp$, and $U$ denotes the on-site
Coulomb repulsion.
We assume the strong coupling limit, $U/t,U/t_\perp \gg 1$ and start out with
the case of half-filling and vanishing leg-hopping $t$.
The system then decomposes into an array of Heitler-London type
dimers and the gound state
is simply the product of singlets along the rungs of the ladder,
with energy $E_G \approx -2Nt_\perp^2/U$, where $N/2$ is the
number of rungs. 
%%%%%%%%%%%%%%%%%%%%%%%%%%%%%%%%%%%%%%%%%%%%%%%%%%%%%%%%
\begin{figure}
\epsfxsize=6cm
\hspace{1cm}\epsffile{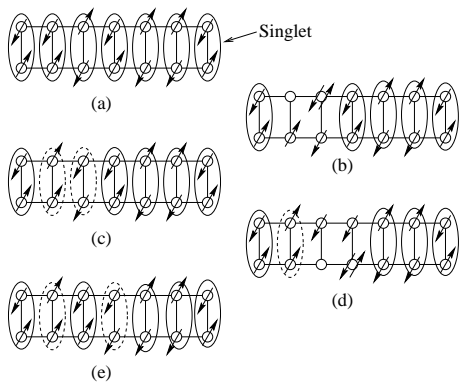}
\vspace{0.5cm}
\narrowtext
\caption{Starting from the RVB vacuum (a)
the `virtual' hopping process
(a) $\rightarrow$ (b) $\rightarrow$ (c) leads to the
pair creation of two triplets (indicated by dashed lines).
Further virtual hopping processes such as
(c) $\rightarrow$ (d) $\rightarrow$ (e) enable the
propagation of the triplets along the ladder}
\label{fig1}
\end{figure}
%%%%%%%%%%%%%%%%%%%%%%%%%%%%%%%%%%%%
\noindent
This state is frequently referred to as the
`rung RVB state' and has been the basis
of many theoretical works on spin ladders\cite{Gopalan,ladder,bulk}.
Next, let us consider the effect of switching on
$t$. This will produce charge fluctuations, and hence
exchange processes along the
legs. These lead, as a first 
step, to the `pair creation' of triplets
along the legs. In
subsequent steps, these newly created triplets
can propagate along the ladder (see Figure \ref{fig1}).
Introducing operators which create singlets
and triplets\cite{Sachdev}:
\begin{eqnarray}
s_{ij}^\dagger &=& \frac{1}{\sqrt{2}}
(\hat{c}_{i,\uparrow}^\dagger \hat{c}_{j,\downarrow}^\dagger
- \hat{c}_{i,\downarrow}^\dagger \hat{c}_{j,\uparrow}^\dagger),
\nonumber \\
t_{ij,x}^\dagger &=& 
-\frac{1}{\sqrt{2}}(
\hat{c}_{i,\uparrow}^\dagger \hat{c}_{j,\uparrow}^\dagger
-\hat{c}_{i,\downarrow}^\dagger \hat{c}_{j,\downarrow}^\dagger),
\nonumber \\
t_{ij,y}^\dagger &=& 
\frac{i}{\sqrt{2}}(
\hat{c}_{i,\uparrow}^\dagger \hat{c}_{j,\uparrow}^\dagger +
\hat{c}_{i,\downarrow}^\dagger \hat{c}_{j,\downarrow}^\dagger),
\nonumber \\
t_{ij,z}^\dagger &=& \frac{1}{\sqrt{2}}
(\hat{c}_{i,\uparrow}^\dagger \hat{c}_{j,\downarrow}^\dagger
+ \hat{c}_{i,\downarrow}^\dagger \hat{c}_{j,\uparrow}^\dagger),
\end{eqnarray}
we can write the rung-RVB state
(which we henceforth consider as a kind of `vacuum'), as
\begin{equation}
|\Omega\rangle = \prod_{n=1}^{N/2}
s_{n}^\dagger |0\rangle.
\end{equation}
Here, the site indices along a
rung, $(i,j)$, have been replaced by a single index $n$ labelling the
rung.
It has been shown
by Gopalan {\em et al.}\cite{Gopalan} that the dynamics
of the triplets
can be mapped {\em exactly} onto a system of three species of hard-core
bond-Bosons on a 1D chain. 
Thereby the presence of the Boson $t_a^\dagger$ on some site
means that the corresponding rung is in the triplet $a$ state -
absence of any Boson implies that the rung is in the singlet state.
The $3$ components
of the triplet on the $n^{th}$ rung form a $3$-dimensional vector 
\begin{equation}
\bbox{\tau}_n^\dagger 
= \left(
\begin{array}{c}
t_{n,x}^\dagger\\
t_{n,y}^\dagger\\
t_{n,z}^\dagger\\
\end{array} \right)
\end{equation}
and the Hamiltonian operator governing the triplet-like Bosons
can be cast into the following manifestly $SO(3)$ invariant
form\cite{Gopalan,ladder}:
\begin{eqnarray}
H &=& 
 J_\perp \sum_{ n }
\bbox{\tau}_{n}^\dagger \cdot \bbox{\tau}_{n}^{}
\nonumber \\
&+&  \frac{J}{2} \sum_{ n }
(\;\bbox{\tau}_{n}^\dagger \cdot \bbox{\tau}_{n+1}^\dagger + H.c.\;)
+  \frac{J}{2} \sum_{ n }
(\; \bbox{\tau}_{n}^\dagger \cdot \bbox{\tau}_{n+1}^{} + H.c.\;) 
\nonumber \\
&-&\frac{J}{2}
\sum_n :(\;
\bbox{\tau}_n^\dagger \cdot \bbox{\tau}_{n+1}^\dagger\;
\bbox{\tau}_{n+1}^{} \cdot \bbox{\tau}_{n}^{}
\nonumber \\
&\;&\;\;\;\;\;\;\;\;\;\;\;\;\;\;\;\;\;\;\;\;\;\;\;\;\;\;\;
-\bbox{\tau}_n^\dagger \cdot \bbox{\tau}_{n+1}^{}\;
\bbox{\tau}_{n+1}^\dagger \cdot \bbox{\tau}_{n}^{}\;):
\label{triplets}
\end{eqnarray}
Here $::$ as usually denotes normal ordering and
the parameters are $J$$=$$4t^2/U$ and $J_\perp$$=$$4t_\perp^2/U$.
The terms in (\ref{triplets}) describe 
the `energy of formation' of a triplet (1st term);
the pair creation of two triplets on neighboring
rungs (2nd term); the propagation of a triplet (3rd term),
the simultaneous `species flop' of two triplets on
neighboring rungs (4th term); and the exchange
of two triplets (5th term)\cite{Gopalan,ladder}.\\
Let us now consider a new type of fluctuation.
Once a charge fluctuation has been created
one might envisage that the remaining electron
in the singly occupied rung follows suit, so that the
resulting state has one empty rung, and a fourfold occupied rung (see
Figure \ref{fig2}). Again,
%%%%%%%%%%%%%%%%%%%%%%%%%%%%%%%%%%%%%%%%%%%%%%%%%%%%%%%%
\begin{figure}
\epsfxsize=6cm
\hspace{1cm}\epsffile{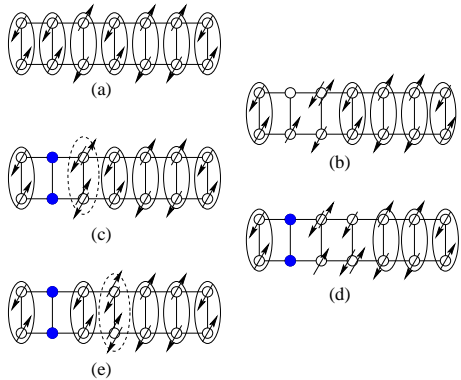}
\vspace{0.5cm}
\narrowtext
\caption{A second type of fluctuation which leads to the creation
of charged rungs: starting from the vaccum,
(a) $\rightarrow$ (b) $\rightarrow$ (c) leads to the
creation of an empty and a fourfold occupied rung.
Further virtual hopping processes such as
(c) $\rightarrow$ (d) $\rightarrow$ (e) enable the
propagation of these charged excitations.}
\label{fig2}
\end{figure}
%%%%%%%%%%%%%%%%%%%%%%%%%%%%%%%%%%%%
\noindent
in subsequent steps the empty and fourfold
occupied rung can propagate, in much the same way
as the triplets did.
Clearly, the fourfold occupied rung has a very high energy $\approx 2U$.
Following Ref.\cite{S-Z-H} however, we now amend the Hamiltonian
with the goal to `pull down' this state in energy,
so as to make the empty-fourfold state degenerate with the
triplet-triplet state.
This can be achieved by adding terms of the form
\begin{equation}
H_1 = \sum_{i,j} V_{i,j} (n_i-1) (n_j-1) +  \sum_{i,j} J_{i,j} \bbox{S}_i,
\cdot \bbox{S}_j,
\label{hcorr}
\end{equation}
where
$n_i= \sum_\sigma c_{i,\sigma}^\dagger c_{i,\sigma}$ is the
operator of electron density on site $i$, and
$\bbox{S}_i$ the operator of electron spin.
It has been shown in Ref.\cite{S-Z-H} that by retaining
only a density interaction $V$ and exchange constant
$J$ within the rungs, and choosing $U+V$$=$$J/4$, one can reach a situation
where the empty-fourfold fluctuation in Figure \ref{fig2}c is
precisely degenerate with the triplet-triplet
fluctuation in Figure \ref{fig1}c, and both of them are lower than 
the singly-threefold intermediate state
(Figure\ref{fig1}b, Figure\ref{fig2}b).
In the following, we call the energy of the 
singlet $E_0$, that of the triplet/empty/fourfold occupied rung
$E_1$ and that of a singly/threefold occupied rung $E_2$.
The situation we want to reach is $E_0 < E_1 \ll E_2$.
We would like to stress that the parameters we need to choose to
reach this ($U+V$$=$$J/4$) are rather unphysical.
Physically, if we want $J$ and $U$ to be positive
and $J$ not to be too large, this requires a large negative rung 
interaction $V$.
However, for the time being we
ignore this complication and defer a discussion
of the case $H_1 \rightarrow 0$ to the end of this section.\\
If we restrict the Hilbert
space to only rung triplets and singlets, the Hamiltonian 
(\ref{triplets}) remains
valid also for the $SO(5)$ symmetric ladder. The only
difference is a change of $J$ to $J'=4t^2/(E_2-E_1)$
and a change of the Boson's `energy of formation' from
$J_\perp \rightarrow \Delta_1=E_1-E_0$. Our goal is now to
simply retain this Hamiltonian, but enlarge the
3-dimensional vector $\bbox{\tau}$ into a 5-dimensional vector comprising 
2 additional
Bosons which represent the empty and fourfold occupied rung. 
We introduce new Bosons $h$ and $d$, whose presence
on a rung implies that the rung is empty or fourfold occupied:
\begin{eqnarray}
h_{i,j}^\dagger &\rightarrow& |vac\rangle \nonumber \\
d_{i,j}^\dagger &\rightarrow& - 
c_{j,\uparrow}^\dagger
c_{i,\uparrow}^\dagger
c_{j,\downarrow}^\dagger
c_{i,\downarrow}^\dagger | vac \rangle
\end{eqnarray}
(the extra minus sign makes sure that the state corresponding to
$d_{i,j}^\dagger$ can be written as
$\Delta_{i,j}^\dagger s_{i,j}^\dagger$, with $\Delta$ the singlet-pairing
operator along a rung).
Just as two singlets on neighboring rungs can
convert themselves in a kind of pair creation
process (Figure\ref{fig1}a $\rightarrow$ Figure\ref{fig1}c) into two triplets,
they can also convert themselves into an empty rung
and a fourfold occupied rung. And, similarly as
a triplet can exchange itself with a singlet,
so can an empty or doubly occupied rung.
An analogous process to the 4th term would be
the conversion of two triplets on neighboring rungs into
an empty and a fourfold occupied rung.
And finally, an empty rung can exchange itself also with
a triplet, which corresponds to the 5th term.
in Eqn.(5).\\
The empty rungs and fourfold occupied rungs thus
will act `almost' like the triplets - there is, however,
a subtle and crucial difference. Careful
calculation shows that all matrix elements for
pair creation and propagation of the empty and fourfold
occupied rungs have the same magnitude, but
{\em opposite sign} as for the triplet rungs.
The ultimate reason is that a spin-flip (which propagates a
triplet) is accomplished by a back-and-forth
motion of electrons, and hence picks up a Fermi minus
sign, whereas the propagation of an empty rung corresponds to
a net movement of charge in one direction,
and hence gets no Fermion minus sign.
We cope with this by a rung-dependent gauge
transformation for the charged Bosons and introduce:
\begin{eqnarray}
t_{n,1}^\dagger &=&
e^{i\pi n} \frac{1}{\sqrt{2}}
( d_{n}^\dagger + h_{n}^\dagger ),
\nonumber \\
t_{n,5}^\dagger &=&
e^{i\pi n} \frac{i}{\sqrt{2}}
( d_{n}^\dagger - h_{n}^\dagger ).
\label{bosons}
\end{eqnarray}
The extra phasefactor $e^{i\pi n}$ precisely cancels
the minus sign in the matrix elements of the two new
Bosons, so that we can now describe the
physics of the Bosons by the following manifestly $SO(5)$ invariant
Hamiltonian
\begin{eqnarray}
H &=& 
 \Delta_1 \sum_{ n }
\bbox{t}_{n}^\dagger \cdot \bbox{t}_{n}^{}
\nonumber \\
&+&  \frac{J'}{2} \sum_{ n }
(\;\bbox{t}_{n}^\dagger \cdot \bbox{t}_{n+1}^\dagger + H.c.\;)
+  \frac{J'}{2} \sum_{ n }
(\; \bbox{t}_{n}^\dagger \cdot \bbox{t}_{n+1}^{} + H.c.\;) 
\nonumber \\
&-&\frac{J'}{2}
\sum_n :(\;
\bbox{t}_n^\dagger \cdot \bbox{t}_{n+1}^\dagger\;
\bbox{t}_{n+1}^{} \cdot \bbox{t}_{n}^{}
-\bbox{t}_n^\dagger \cdot \bbox{t}_{n+1}^{}\;
\bbox{t}_{n+1}^\dagger \cdot \bbox{t}_{n}^{}\;):
\label{heff1}
\end{eqnarray}
Here $\bbox{t}$ denotes the 5-dimensional vector
\begin{equation}
\bbox{t}_n^\dagger = \left(
\begin{array}{c}
t_{n,1}^\dagger\\
t_{n,x}^\dagger\\
t_{n,y}^\dagger\\
t_{n,z}^\dagger\\
t_{n,5}^\dagger\\
\end{array} \right).
\end{equation}
In terms of this $5$-dimensional Boson-vector on a single rung
the $10$ root generators of $SO(5)$ take the simple form
(suppressing the rung index)
\begin{equation}
L_{ab} = -i ( t_a^\dagger t_b^{} - t_b^\dagger t_a^{} ),
\label{root}
\end{equation}
which for $x \leq a,b \leq z$ reduces to a representation of the
$SO(3)$ spin operators in the rung basis\cite{ladder}.
Bearing in mind that the $\bbox{t}$ are hard-core
Bosons, it is straightforward to show
that the operators (\ref{root})
obey the $SO(5)$ angular momentum algebra\cite{science}:
\begin{equation}
[ L_{ab}, L_{cd} ] = -i(
\delta_{ad} L_{bc}
-\delta_{ac} L_{bd}
+ \delta_{bc} L_{ad}
-\delta_{bd} L_{ac} ).
\label{so5_algebra}
\end{equation}
Choosing $\alpha \in x,y,z$, the combination
\begin{equation}
\pi_\alpha = \frac{1}{\sqrt{2}}(L_{1\alpha} - i L_{\alpha 5})
\end{equation}
replaces an $a$-type triplet with
momentum $k=\pi$ by a hole pair along the rungs with
momentum $k=0$  -
this operator is therefore nothing but the ladder equivalent of the
$\pi$-operator in 2D\cite{science}, which replaces
a triplet with momentum $(\pi,\pi)$ by a $d_{x^2-y^2}$ hole pair
with momentum $(0,0)$.\\
We can also express the $SO(5)$ hard-core Boson Hamiltonian (\ref{heff1})
directly in terms of a $SO(5)$ quantum nonlinear $\sigma$ model Hamiltonian by
introducing the $SO(5)$ superspin vector at a given rung as  
\begin{equation}
x_a = \frac{1}{\sqrt{2}} (t_a + t_a^\dagger)   
\end{equation}
and its conjugate momentum as
\begin{equation}
p_a = \frac{1}{i\sqrt{2}} (t_a - t_a^\dagger).
\end{equation}
In this representation, the symmetry generator takes the more familiar
form
\begin{equation}
L_{ab} = x_a p_b - x_b p_a
\end{equation}
Using these operator identities, the $SO(5)$ hard-core Boson 
Hamiltonian (\ref{heff1}) can be expressed as
(repeated $SO(5)$ indices are summed over!)
\begin{eqnarray}
H &=& 
 \frac{\Delta_1}{4} \sum_{ n } L_{ab}^2(n) 
+ J' \sum_{ n } x_a(n) x_a(n+1)
\nonumber \\
&+&  \frac{J'}{4} \sum_{ n }
L_{ab}(n) L_{ab}(n+1)
\label{sigma}
\end{eqnarray}
This Hamiltonian is quantized using the $SO(5)$ commutation relations
(\ref{so5_algebra}) and 
\begin{equation}
[ L_{ab}, x_c ] = -i(
\delta_{bc} x_a - \delta_{ac} x_b) ,
\end{equation}
together with the hard core constraint
\begin{equation}
x_a x_b = \delta_{ab}.
\end{equation}
The $SO(5)$ ladder model can be used for 
a particularly simple and transparent demonstration
of the key feature of the $SO(5)$ theory, namely the
one-to-one correspondence of antiferromagnetism
and superconductivity. The ground state of the ladder
models is actually an RVB type of vacuum without
AF long-range-order. However, for
illustrative purposes, let us now construct an AF ordered
state (which is in general not a eigenstate of the
Hamiltonian) by condensing the magnons into the
RVB ground state. 
We can obviously express the operator of staggered magnetization
in $z$-direction as
\[
M_s = \sum_n e^{i \pi n}
(\;P_n(\uparrow \downarrow) - P_n(\downarrow \uparrow)\;),
\]
where e.g. $P_n(\uparrow \downarrow)$ projects onto states
where the $n^{th}$ rung has the configuration $\uparrow \downarrow$.
It is now easy to see that
\begin{eqnarray}
(\;P_n(\uparrow \downarrow) - P_n(\downarrow \uparrow)\;)
s_n^\dagger &=& t_{n,z}^\dagger, \nonumber \\
(\;P_n(\uparrow \downarrow) - P_n(\downarrow \uparrow)\;)
t_{n,a}^\dagger &=& \delta_{a,z}\; s_n^\dagger,
\end{eqnarray}
whence 
\[
M_s = \sqrt{\frac{N}{2}}[\;
 t_z^\dagger(q=\pi) + t_z^{}(q=\pi)\;].
\]
If we now form the coherent state
\[
|\Psi_\lambda \rangle =
\frac{1}{\sqrt{n}} e^{ \lambda \sqrt{N}
t_z^\dagger(q=\pi)} |\Omega\rangle,
\]
which corresponds to $z$-like triplets condensed into the
$k$$=$$\pi$ state, and
treat the $\bbox{t}$ as ordinary Bosons, we obtain
\[
\langle \Psi_\lambda | M_s | \Psi_\lambda \rangle =
\sqrt{2} \lambda N.
\] 
If the hard-core constraint is taken into account
rigorously, the only change is 
an extra correction factor of $1/(1+\lambda^2)$ on the r.h.s.,
see the Appendix. \\
This calculation shows that by starting from an `RVB vacuum',
an antiferromagnetically ordered state with $M_S$ in $z$-direction
can be generated by
condensing $z$-like triplet-excitations into the $k$$=$$\pi$ state.
At this point we can invoke the $SO(5)$ symmetry of the model, 
which tells us that since the  $z$-like triplet with momentum $\pi$ 
and the hole pair with momentum $0$
are two different components of a 5-vector
(the difference in momenta is precisely absorbed by
the gauge transformation (\ref{bosons})
we needed to make the signs of
hopping integrals consistent), they are
dynamically indistinguishable. This means that the
AF state, with condensed triplets, can be $SO(5)$-rotated
into a state with condensed hole pairs.
It follows that if the antiferromagnetic state were
the ground state at half-filling (which is the case for 2D materials)
we can replace all $z$-like triplets by hole pairs
with momentum $0$ and by $SO(5)$ symmetry automatically
obtain the ground state in the doped case. The latter then
consists of hole-pairs along the rungs condensed into the
$k=0$ states and thus is necessarily superconducting.
In other words: both the antiferromagnetic and the
superconducting state may be viewed as some kind of
condensate `on top of' the rung-RVB state.
$SO(5)$ symmetry then simply implies that the
condensed objects are combined into a single
vector, whence the unification of antiferromagnetism and
superconductivity follows in a most natural way.\\
The above derivation makes sense only in a strong coupling limit,
where a discussion starting out from rung-singlets makes sense.
One might expect, however, that similar considerations will apply
also for cases with a weak coupling within the 
rungs\cite{Enrico,Fisher}.\\
To conclude this section, we discuss what will happen if we
switch off the correction terms (\ref{hcorr}), 
which were introduced so as to
enforce exact $SO(5)$ symmetry. The crucial question is whether
above considerations will retain some validity or break down
alltogether. Let us consider the $1-5$ plane of the $5$-dimensional
space, i.e. the `charge-like subspace'.
In the fully $SO(5)$-symmetric model the vector $\bbox{t}$ can be
rotated freely in this plane.
The total charge is related to the angular momentum in the 
$1-5$ plane, since $Q=L_{15}=x_1 p_5 - x_5 p_1$. 
For a system with a large charge gap compared with 
the spin gap, one can apply a chemical potential to lower the
energy of the empty rungs $h_n^\dagger|0\rangle$ at the cost of
further increasing the energy of the fourfold occupied rungs
$d_n^\dagger|0\rangle$. The chemical potential
acts like a magnetic field normal to the $1-5$ plane, and selects
a particular sense of the rotation in the $1-5$ plane.
Therefore, the ``free rotor" in the exact $SO(5)$ symmetric models
becomes a ``chiral rotor" in the more realistic Hubbard or $t-J$ models.\\
In terms of the Boson model, switching off $H_1$
will lead to different energies for the different Boson species, i.e.
a replacement of the form
\[
 \Delta_1 \sum_{ n }
\bbox{t}_{n}^\dagger \cdot \bbox{t}_{n}^{}
\rightarrow \sum_n \epsilon_0 h_{n}^\dagger  h_{n}
+ J \bbox{\tau}_n^\dagger \cdot \bbox{\tau}_n
+ \epsilon_2  d_{n}^\dagger  d_{n},
\]
where $\epsilon_0$ corresponds to the binding energy of a hole pair
(relative to the rung singlet) and $\epsilon_2\approx 2U$.
Here we are neglecting the modification of the
off-diagonal matrix elements $\propto J'$.
While $\epsilon_2$ is a huge energy, let us assume that
we have a state with $Q\leq 0$ 
and treat this term in
perturbation theory. The crucial point is now, that
for total electron density $\leq 1$ the fourfold
occupied rung $ d_{n}^\dagger$ is admixed only as a quantum
fluctuation, so that we expect for the change in energy
\[
\delta E =  (\epsilon_0-\Delta_1) Q +  
0(\left( \frac{J'}{E_1-E_0} \right)^2).
\]
In the limit $J' \ll E_1-E_0$ the main effect of 
switching to the physical Hubbard or $t-J$ model {\em in the
hole-doped subspace} thus is adding a chemical potential-like term
$(\epsilon_0-\Delta_1) Q$ to the Hamiltonian. Apart from that, having
$\epsilon_2 \rightarrow 2U$ for states with $Q\leq0$ merely
corresponds to projecting out quantum fluctuations
involving charged Bosons. This may be expected to have only a
minor effect. Of course states with $Q>1$ will
have their energies shifted by $Q\epsilon_2$, so that
$SO(5)$ rotations into this sector of the Hilbert space become
forbidden. As long as we restrict ourselves to
hole-doped states, however, one may hope that
$SO(5)$ symmetry remains approximately valid.
Below we will present an explicit numerical check of this more
qualitative discussion.
%%%%%%%%%%%%%%%%%%%%%%%%%%%%%%%%%%%%%%%%%%%%%%%%%%%%
\section{Classification of states according to SO(5) multiplets}
\begin{figure}
\epsfxsize=7.5cm
\hspace{0.5cm}\epsffile{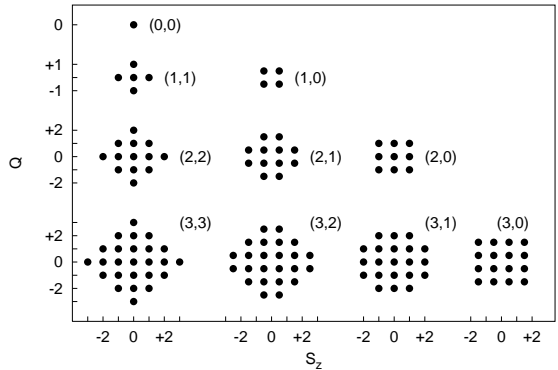}
\narrowtext
\caption{The first few irreps of $SO(5)$ displayed on the
$Q-S_z$-plane. The $(p,p)$ multiplets have the familiar
diamond shape; with $(p,q<p)$ the multiplets evolve to a
square shape. The number of dots in each multiplet is not the
full dimension of the $SO(5)$-irrep since there are additional
degeneracies in the $S_z$-direction.}
\label{fig3}
\end{figure}

\begin{figure}
\epsfxsize=7.5cm
\epsffile{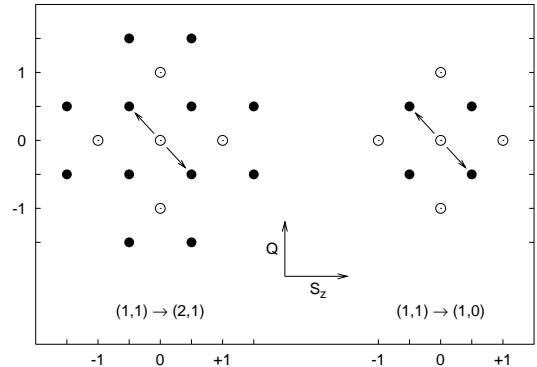}
\hspace{1cm}\narrowtext
\caption{Visualization of the selection rules for
photoemission and inverse photoemission for $SO(5)$ symmetric
states on the example of a transition originating from a half filled
state with $S_z=0$ in the $(1,1)$ multiplet. 
$SO(5)$ selection rules only allow
transitions to the $(1,0)$ and the $(2,1)$ multiplets (black
discs), which
are superimposed on the initial $(1,1)$ multiplet (circles). In this
example, the photoemission and inverse photoemisson process
remove/inject a spindown-electron so the photoemission
transition is visualized by an arrow pointing south-east and
the inverse photoemission arrow points towards north-west.
}
\label{fig4}
\end{figure}

\begin{figure}
\epsfxsize=7.5cm
\epsffile{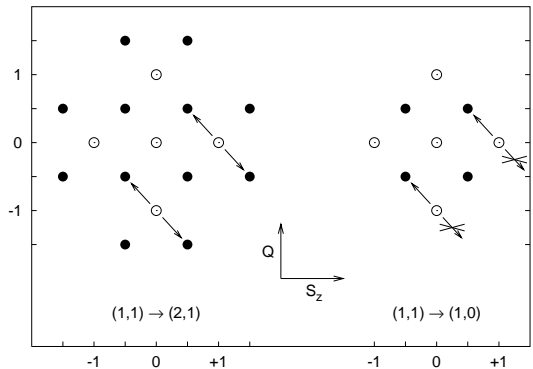}
\narrowtext
\caption{Photoemission and inverse photoemission originating
from a spin-polarized half-filled state ($Q=0, S_z=+1$) and from
a doped state ($Q=-1, S_z=0$). Both states are members of the
$(1,1)$ multiplet like the initial state in Figure 4.
%SCZ%\ref{fig5}. 
This Figure shows that there is no allowed
photoemission transition to the $(1,0)$ multiplet
for either of the two states.
}
\label{fig5}
\end{figure}

%%%%%%%%%%%%%%%%%%%%%%%%%%%%%%%%%%%%%%%%%%%%%%%%%%%%
In this section we shall briefly review the $SO(5)$ group theory and
classify all states in the ladder model according to irreducible
representations (irreps) of the $SO(5)$ Lie algebra. We also 
discuss various $SO(5)$ selection rules which shall be used in 
the next two sections.\\
In reference \cite{ede97}, three of us showed how low lying bosonic
states in the 
$t-J$ model can be classified by the fully symmetric tensor multiplets
of the $SO(5)$ Lie algebra. The full multiplet structure of the 
$SO(5)$ group is much richer. 
The general irreps of an
$SO(5)$-symmetric model are described by two integers 
\begin{equation}
(p,q), \, p \geq q \geq 0 
\label{genrep}
\end{equation}
with dimension
\begin{equation}
D = \left(1+q\right) \left(1+p-q\right) \left(1+\frac{1}{2} p\right) 
\left(1+\frac{1}{3} 
\left(p+q\right)\right) \, ,
\end{equation}
and Casimir\cite{hu}
\begin{equation}
C = \frac{1}{2} p^2 + \frac{1}{2} q^2 + q + 2 p .
\label{c1}
\end{equation}
$SO(5)$ is a rank two algebra so we choose the charge $Q$
and
the z-component of the Spin $S_z$ as members of the Cartan 
subalgebra of mutually commuting generators. This allows us to
draw the different irreps in the $Q-S_z$ plane. The $(p,p)$
irreps have the familiar \cite{ede97} diamond shape and corresponds to
the fully symmetric traceless tensor irreps identified in \cite{ede97}.
The $(p,0)$ irreps form a square in the $Q-S_z$ plane and are the
spinor states found in systems with odd number of electrons. 
For example, the quintet $E_1$ manifold with 3 magnons and 2 pair
states carry the label $(1,1)$, while the quartet $E_2$ manifold with
one or three electrons per rung carries the label $(1,0)$.   The
intermediate multiplets $(p,p > q > 0)$ are the
evolution from the square to the diamond (Figure \ref{fig3}).
All states of the exact $SO(5)$ symmetric models 
%%%%%%%%%%%%%%%%%%%%%%%%%%%%%%%%%%%%%%%%%%%%%%%%%%%%%%%%
%%%%%%%%%%%%%%%%%%%%%%%%%%%%%%%%%%%%
\noindent
constructed in \cite{S-Z-H} can be classified into this $SO(5)$ multiplet 
structure. We have verified this numerically by computing the energies and
expectation values of the Casimir operator for 
{\em all} eigenstates of a $2\times 4$ system. 
By scanning all sectors of different hole number and $z$-spin
and collecting states of equal (within computer accuracy, i.e. $10^{-12}$) 
energy and Casimir charge 
we have verified that from the ground states
up to the highest excited states all eigenstates of the system can be
classified into the multiplets shown in Figure \ref{fig3}
(and more complicated ones). For the $2\times 6$ system, where
a full digonalization is not feasible anymore, we have
verified this for the low energy states obtainable by Lanczos
diagonalization.\\
Besides classifying eigenstates into $SO(5)$ multiplet structures,
$SO(5)$ symmetry also gives powerful selection rules on the possible
tranition processes and relates various matrix elements through the
Wigner-Eckart theorem. In the following two sections, we are interested
in the photoemission, charge and spin spectra of the $SO(5)$ symmetric
ladder model. The perturbing operator in the photoemission process
is a single electron operator, which transforms according to the 
4 dimensional $(1,0)$ irreps under $SO(5)$. In this work, we shall 
consider the initial and final states of the form $(p,p)$, corresponding
to excited magnon and pair states. The possible final states
generated by the perturbing operator can be obtained by decomposing the
product representation into irreducible components, and it given by
\begin{equation}
(1,0) \otimes (p,p) = (p+1,p) + (p,p-1) 
\end{equation}
Similarly, the perturbing operator in the neutron scattering or
Josephson tunneling process transform according to the 5 dimensional
$(1,1)$ irreps under $SO(5)$. The possible final states are
given by the following decomposition rules:
\begin{eqnarray}
(1,1) \otimes (p,p) &=& (p+1,p+1) 
\nonumber \\
&\;&\;\;\;\;\;+ (p+1,p-1) + (p-1,p-1) 
\end{eqnarray}
The photoemission selection rules (except for 
$S \longrightarrow S \pm \frac{1}{2}$)
can be easily visualized by superimposing the involved
$SO(5)$-multiplets (Figures \ref{fig4} and
\ref{fig5}). 
Figure \ref{fig4} shows that there are four possible
photoemission and inverse photoemission transitions for the
$S_z=0$ state at half-filling which belongs to the $(1,1)$
multiplet. We 
%%%%%%%%%%%%%%%%%%%%%%%%%%%%%%%%%%%%%%%%%%%%%%%%%%%%%%%
%%%%%%%%%%%%%%%%%%%%%%%%%%%%%%%%%%%%%%%%%%%%%%%%%%%
\noindent
therefore expect four distinct bands in the
spectrum. If, on the other hand, one takes a spin-polarized or
doped state of the same multiplet as the initial state for
photoemssion and inverse photoemission we expect to see only
three bands since there are only three possible transitions
according to the selection rules (Figure \ref{fig5}).
Furthermore, we expect the {\it very same} spectrum for both 
the polarized and the doped case
since the initial states belong to the same
initial multiplet as well as all target states belong to the same
target multiplets and are therefore degenerate. 
%%%%%%%%%%%%%%%%%%%%%%%%%%%%%%%%%%%%%%%%%%%%%%%%
%%%%%%%%%%%%%%%%%%%%%%%%%%%%%%%%%%%%%%%%%%%%%%%%%%%%%%%%
\section{Single particle spectra}
\begin{figure}
\epsfxsize=8cm
\vspace{-0.5cm}
\epsffile{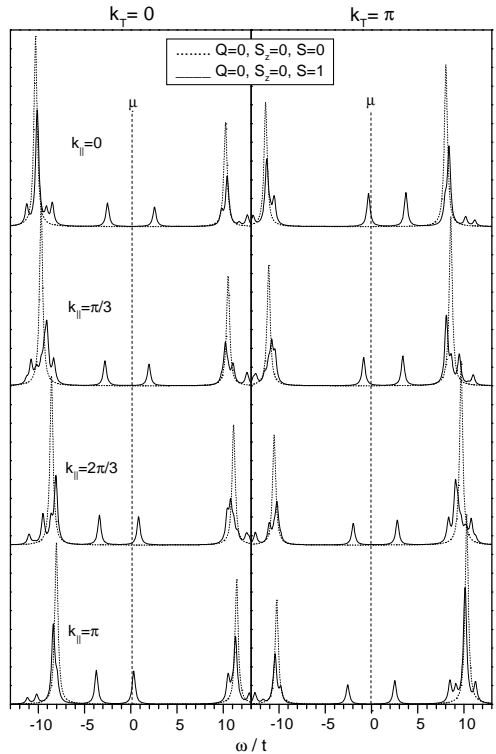}
\vspace{-0.0cm}
\narrowtext
\caption{Comparison of the single particle spectral function for
the half-filled ground state (dotted line) and the half-filled
$^3(\pi,\pi)_0$ state with $S_z$$=$$0$ (full line). 
The Fermi energy, defined as the
average of first ionization and affinity energy, is taken as
the zero of energy. The parameter values are
$t=1$, $t_\perp=1$, $U=8$, $V=-6$, and $J=8$.}
\label{fig6}
\end{figure}

%%%%%%%%%%%%%%%%%%%%%%%%%%%%%%%%%%%%%%%%%%%%%%%%%%%%%%%%
We now want to see the practical application of the
selection rules discussed in the preceeding section. To that end
we study the photoemission (PES) and
inverse photoemission (IPES) spectrum, which we define as
\begin{eqnarray*}
A_{PES}(\bbox{k},\omega) &=& \frac{1}{\pi} \Im
\langle 0 |c_{\bbox{k}\downarrow}^\dagger\frac{1}{
\omega +H-\epsilon_0-i0^+}c_{\bbox{k}\downarrow}^{}| 0 \rangle,
\nonumber \\
A_{IPES}(\bbox{k},\omega) &=& \frac{1}{\pi} \Im
\langle 0 |c_{\bbox{k}\downarrow }^{}\frac{1}{
\omega -H+\epsilon_0-i0^+}c_{\bbox{k}\downarrow}^\dagger| 0 \rangle.
\end{eqnarray*}
Here $|0\rangle$ is a suitably chosen initial state and $\epsilon_0$ 
its energy.
In the following, we will label different states
as $^m\bbox{K}_Q$, where $m$ denotes the standard
spin multiplicity, $\bbox{K}$ the total momentum and
$Q$ the charge quantum number. For finite clusters
of our $SO(5)$ symmetric ladder model this is readily
obtained numerically by Lanczos diagonalization\cite{Dagoreview}.
The $SO(5)$ multiplets discussed above are
easily identified, because the energies of states
belonging to one multiplet are degenerate (i.e. identical
to computer accuracy). This allows to study the
evolution of the single-particle spectral function
as we pass from one multiplet ($p,q$) to the other, and 
within one multiplet through the different doping levels.
To begin with, Figure \ref{fig6} shows the single-particle spectrum
for the half-filled ground state $^1(0,0)_{0}$, which actually forms a
one-dimensional $(0,0)$ multiplet. Final states can only
belong to the $4$-dimensional $(1,0)$ representation,
see Figure \ref{fig3}. Despite
the fact that we are using very strong interaction parameters, there
is just one single electron-like
band in PES, whose dispersion closely follows the
noninteracting dispersion.
The center of gravity of this band 
is given by the energy difference between a rung-singlet and a
singly occupied rung, i.e. $E_0-E_2=-7U/2-3V$, i.e. $-10$ 
with our present parameter values. We proceed to the spectrum 
of the $^3(\pi,\pi)_0$ state, with $S_z$$=$$0$ - this state 
%%%%%%%%%%%%%%%%%%%%%%%%%%%%%%%%%%%%%%%%%%%%%%%%%%%%%%%%
%%%%%%%%%%%%%%%%%%%%%%%%%%%%%%%%%%%%%%%%%%%%%%%%%%%%%%%%%%%%%%%%
\noindent
belongs  to the fivefold degenerate $(p=1,q=1)$ multiplet.
From Figure \ref{fig4} we expect final states belonging to both
the $(2,1)$ and the $(1,0)$ representation.
It appears as if the bands seen in the ground state spectra
remain practically unchanged - as a new
feature, howewer, there appear some weak `sidebands' close to
$\mu$. They can be seen both in photoemission and in inverse
photoemisson, so that we obviously have precisely the
$4$ bands expected on the basis of our discussion of
selection rules given above (see Figure \ref{fig4}).
The physical interpretation is straightforward.
The $^3(\pi,\pi)_0$ state has a single triplet-like
Boson. The `main bands', which are similar to those
in the ground state spectra,
correspond to final states where the
photohole is generated in a singlet-rung.
They therefore contain the original triplet,
plus a singly occupied rung which propagates through
the ladder and carries the entire momentum
transfer. These states therefore belong to the
$(2,1)$ representation. 
The center of gravity of this band 
is again at $E_0-E_2$. The two excitations (triplet and hole)
now can scatter from each other, whence the respective band becomes
broadened. There is however also a second process, namely the photohole
can be created in the rung occupied by the triplet.
This creates final states containing only a singly occupied rung,
which must therefore belong to the $(1,0)$ representation.
As the intensity of this second process is proportional
to the triplet density, our interpretation of the sidebands
can be checked numerically by comparing the mean weight of the 
sidebands for one-triplet-states on ladders 
with different numbers of rungs. On 4-, 6-, and 8-rung ladders
(which corresponds to triplet densities of $1/4$, $1/6$, and $1/8$)
and with our parameter values, we found mean weights of $0.276$, $0.183$,
and $0.137$ -- a convincing proof of the interpretation given
above. The center of gravity 
%%%%%%%%%%%%%%%%%%%%%%%%%%%%%%%%%%%%%%%%%%%%%%%%%%%%%%%%
\begin{figure}
\epsfxsize=8cm
\vspace{-0.5cm}
\epsffile{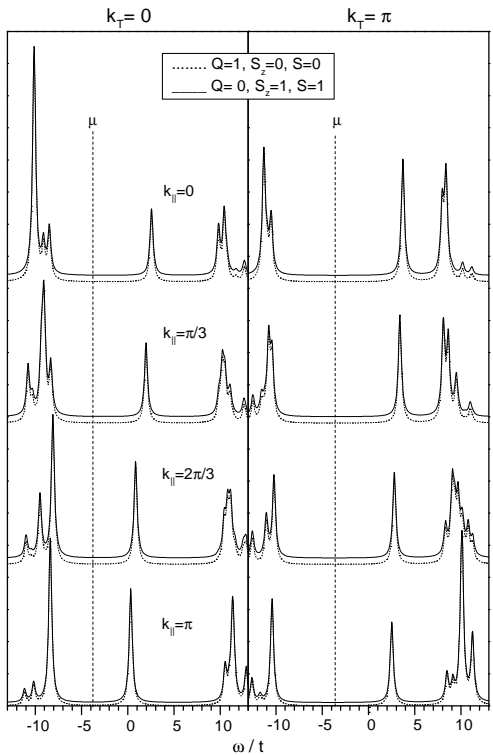}
\narrowtext
\vspace{-0.25cm}
\caption{Comparison of the electron removal spectrum for
the two-hole ground state $^0(0,0)_{-1}$(dotted line) and the half-filled
$^3(\pi,\pi)_0$ state with $S_z$$=$$1$ (full line). 
The removed electron has $\downarrow$-spin. 
The spectra for the half-filled state have been offset in
$y$-direction so as to faciliate the comparison.
Parameter values are as in Figure 6.}
%SCZ%\ref{fig7}
\label{fig7}
\end{figure}
\noindent
%%%%%%%%%%%%%%%%%%%%%%%%%%%%%%%%%%%%%%%%%%%%%%%%%%%%%%%%%%%%%%%%
of the resulting band is at
the energy difference between a rung triplet and a singly occupied rung,
$E_1-E_2=U/2+V$, i.e. $-2$ with our parameters.
In this type of process
the photohole has to absorb also the
momentum of the triplet, $(\pi,\pi)$, so that the {\em dispersion}
of this band is precisely the same as that of the `main band'
seen for the $^1(0,0)_0$ ground state,
but shifted by $(\pi,\pi)$ (note however the rigid
shift of the band by $E_1-E_0$ due to the difference in
initial state energy).
Literally the same arguments hold for the inverse photoemisison
part, and inspection of Figure \ref{fig6} shows
that all of these features are observed.\\
So far we have seen that spin excitations
in the ground state generate
additional sidebands in the single particle spectrum. At this point,
however, we can invoke the exact $SO(5)$ symmetry of the model,
which tells us that spin polarization and hole/electron
doping are equivalent, in that
the empty rung is the `$SO(5)$ partner' of the triplet rung.
Consequently, Figure \ref{fig7} compares the single particle spectrum
of the $^1(0,0)_{-1}$ and
$^3(\pi,\pi)_0$ states. Both states
belong to the same $(1,1)$ representation -
unambiguous evidence is provided by the fact that their
energies agree to computer accuracy. In Figure \ref{fig7}
we have chosen the $S_z$$=$$1$ member of the triplet, and
study the annihilation of a $\downarrow$-electron. 
>From Figure \ref{fig5} we expect that this simulation probes 
exclusively the $(2,1)$ representation. Since the final 
states for these two different processes belong to the
same $SO(5)$ multiplet, we expect their amplitudes to be directly
related. More precisely we have
$|^3(\pi,\pi)_0\rangle=\pi _{+}^\dagger|^1(0,0)_{-1} \rangle$,
with $\pi_+^\dagger = \pi_x^\dagger + i \pi_y^\dagger$, whence:
\begin{eqnarray*}
&&\langle^3(\pi,\pi)_0,S_z=1 |c_{\bbox{k}\downarrow }^\dagger\frac{1}{
\omega +H-\epsilon_0}c_{\bbox{k}\downarrow }|^3(\pi,\pi)_0,S_z=1 \rangle
\nonumber \\
&&=\langle^1(0,0)_{-1}|\pi_+ c_{\bbox{k}\downarrow }^\dagger\frac{1}
{\omega +H-\epsilon_0}
c_{\bbox{k}\downarrow }\pi_{+}^\dagger|^1(0,0)_{-1} \rangle
\nonumber \\
&&=\langle^1(0,0)_{-1}|
c_{\bbox{k}\downarrow }^\dagger\frac{1}{\omega +H-\epsilon_0}[\pi _{+},\pi
_{+}^\dagger]\; c_{\bbox{k}\downarrow }|^1(0,0)_{-1} \rangle,
\nonumber \\
&&=\langle^1(0,0)_{-1}|
c_{\bbox{k}\downarrow }^\dagger\frac{1}{\omega +H-\epsilon_0}
c_{\bbox{k}\downarrow }|^1(0,0)_{-1} \rangle,
\end{eqnarray*}
Where we have used the fact that the $\pi_+$ operator commutes
with $c_{\bbox{k}\downarrow }$ and the Hamiltonian. Furthermore, it
annihilates the $|^1(0,0)_{-1} \rangle$ state.
Since they belong to the same $SO(5)$ multiplet,
the energies of the $|^1(\pi,\pi)_0\rangle$ and the
$|^0(0,0)_{-1}\rangle$ state are identical, i.e. $\epsilon_0$.
Therefore, as we indeed see in Figure \ref{fig7}, these two
correlation functions are completely identical (within computer
accuracy in the simulation).  The physical reason is that
the photohole cannot be created in the triplet rung
(because the latter contains two $\uparrow$-electrons).
In other words, when `seen through the eyes of
the photohole-operator' the triplet
rung looks like it were `empty', i.e. occupied by holes.
{\it $SO(5)$ symmetry thus implies that the single particle spectra
in the doped ground states actually are {\em identical} to those of
certain states at half-filling}. $SO(5)$ thus presents
a very obvious rationalization for the rigid-band
behavior observed numerically in the single particle spectra of the
2D $t-J$ model\cite{EderOhtaShimozato,Nishimoto}.\\
Next, we return to the photoemission spectrum for the
$S_z$$=$$0$ member of the $^3(\pi,\pi)_0$ state.
In this case, the photohole can be generated
in the triplet rung, and, as discussed above, the
sidebands near $\mu$ appear. The corresponding final states have
only one singly occupied rung remaining in the system. Then, precisely
the same kind 
%%%%%%%%%%%%%%%%%%%%%%%%%%%%%%%%%%%%%%%%%%%%%%%%%%%%%%%%
\begin{figure}
\epsfxsize=8cm
\epsffile{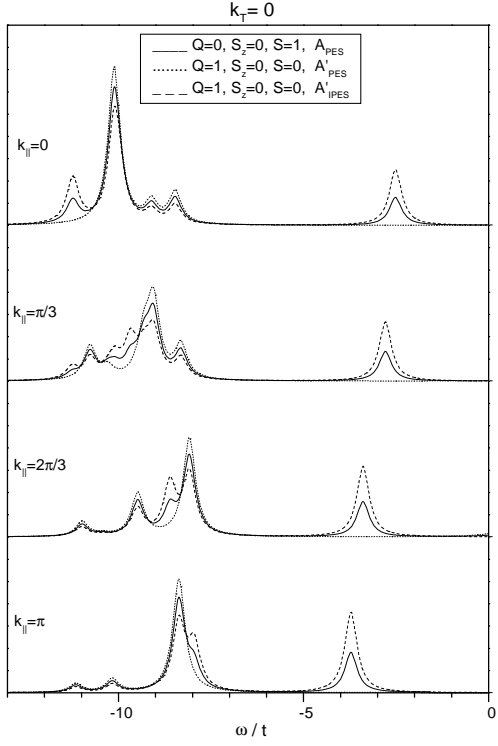}
\narrowtext
\caption{Comparison of the electron removal spectrum for
the two-hole ground state $A_{PES}'(\bbox{k},\omega)$
(dotted line), the `inverted' electron addition spectrum
for the two-hole ground state $A_{IPES}'(-\bbox{k}+\bbox{Q},-\omega)$
(dashed line) and the electron removal spectrum $A_{PES}(\bbox{k},\omega)$
for the $^3(\pi,\pi)_0$ state with $S_z$$=$$0$ (full line). 
Parameter values are as in Figure 6.}
%SCZ%\ref{fig7}
\label{fig8}
\end{figure}
\noindent
%%%%%%%%%%%%%%%%%%%%%%%%%%%%%%%%%%%%%%%%%%%%%%%%%%%%%%%%%%%%%%%%
of state
results if a photoelectron is {\em created} in an empty rung.
The only difference is, that in the former case the
photohole has to absorb the momentum of the triplet
i.e. $(\pi,\pi)$. We can conclude that the sideband seen in PES
at half-filling must disappear
in PES for the two-hole case, but reappear in IPES,
shifted by $(\pi,\pi)$. This can also be seen explicitly
by the following identity:
\begin{eqnarray*}
&&\langle^3(\pi,\pi)_0,S_z=0 |c_{\bbox{k}\downarrow }^\dagger\frac{1}{
\omega +H-\epsilon_0}c_{\bbox{k}\downarrow }|^3(\pi,\pi)_0,S_z=0 \rangle
\nonumber \\
&&=\langle^1(0,0)_{-1}|\pi_z c_{\bbox{k}\downarrow }^\dagger\frac{1}
{\omega +H-\epsilon_0}
c_{\bbox{k}\downarrow }\pi_{z}^\dagger|^1(0,0)_{-1} \rangle
\nonumber \\
&&=\frac{1}{2}\langle^1(0,0)_{-1}|
c_{\bbox{k}\downarrow }^\dagger\frac{1}{\omega +H-\epsilon_0}
c_{\bbox{k}\downarrow }|^1(0,0)_{-1} \rangle,
\nonumber \\
&&+\frac{1}{2}\langle^1(0,0)_{-1}|
c_{-k+Q\uparrow }\frac{1}{\omega +H-\epsilon_0}
c_{-k+Q\uparrow }^\dagger|^1(0,0)_{-1} \rangle,
\end{eqnarray*}
where the second term in the last equation arises from the 
nonvanishing commutator between $\pi_z$ and $c_{\bbox{k}\downarrow}$.
This identity implies that
\begin{equation}
A_{PES}(\bbox{k},\omega)
= \frac{1}{2}
\;[ A_{PES}'(\bbox{k},\omega)
+ A_{IPES}'(-\bbox{k}+\bbox{Q},-\omega)].
\end{equation}
Remarkably enough, the photoemission spectrum $A_{PES}$ of the
half-filled triplet $S_z=0$ state is related to the
photoemission spectrum $A_{PES}'$
and inverse photoemission spectrum $A_{IPES}'$ of
the two hole ground state. Inspection of
Figure \ref{fig8} shows the three spectra in question
demonstrates that the photoemission spectrum for the doped
ground state (full line) indeed is simply the {\it mean} of the two
spectra computed for the doped case.\\
All in all, the preceeding discussion has shown that
$SO(5)$ symmetry leads to at first sight unexpected behavior
in the single-particle spectral function: since spin-polarization
and hole doping are equivalent under $SO(5)$, the
PES spectra of half-filled but spin-polarized states are
identical to those of doped states. This may be the key
to understand the rigid-band behavior observed
\cite{EderOhtaShimozato,Nishimoto} in the 2D $t-J$ model. 
Moreover,
sidebands which appear in PES on spin-excited states at half-filling
and which originate from processes where a spin excitation is
annihilated, reappear in IPES on hole-doped states.
Again, this was found previously also for the 2D $t-J$ model\cite{inverse}.
We note that
the spectroscopies shown in Figures \ref{fig6}, \ref{fig7}, and \ref{fig8}
have been performed for the
actual 2D $t-J$ model as well and have
produced results in strong support of $SO(5)$\cite{tobepub}.
%%%%%%%%%%%%%%%%%%%%%%%%%%%%%%%%%%%%%
\section{Spin and charge dynamics}
\begin{figure}
\vspace{-0.5cm}
\epsfxsize=7cm
\epsffile{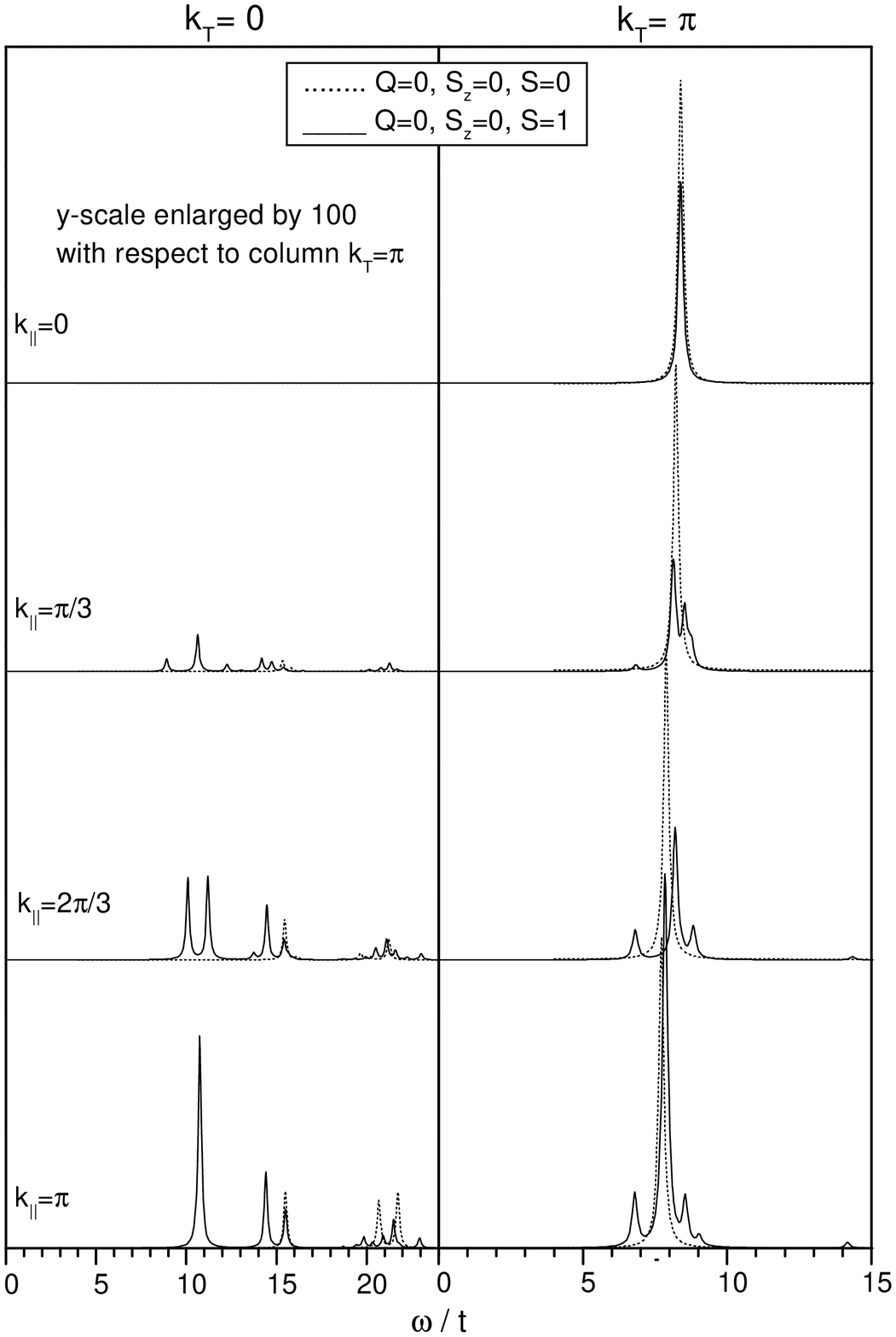}
\narrowtext
\caption{Comparison of the spin correlation spectrum for
the half-filled ground state (dotted line) and the
$^3(\pi,\pi)_0$ state with $S_z$$=$$0$ (full line). 
Parameter values are as in Figure 6.}
%SCZ% \ref{fig6}
\label{fig9}
\end{figure}

%%%%%%%%%%%%%%%%%%%%%%%%%%%%%%%%%%%%%%
Our rung-Boson model gives a description of the physics
in terms of spinful and charged 2-electron excitations,
and thus should be directly applicable to a discussion of the
spin and charge dynamics. We introduce the
spin correlation function (SCF) and the density correlation function
(DCF) as
\begin{eqnarray*}
D_{s}(\bbox{k},\omega) &=& \frac{1}{\pi} \Im
\langle 0 |S_{{-\bf k}}^z\frac{1}{
\omega -H+\epsilon_0-i0^+}S_{\bbox{k}}^z| 0 \rangle,
\nonumber \\
D_{c}(\bbox{k},\omega) &=& \frac{1}{\pi} \Im
\langle 0 |n_{-\bbox{k}}\frac{1}{
\omega -H+\epsilon_0-i0^+}n_{\bbox{k}}| 0 \rangle,
\end{eqnarray*}
where $S^z$ and $n$ denote the operator of
$z$-spin and electron density. As a first step
we need representations
of these operators in terms of the rung-Boson operators
$\bbox{t}$. Following Ref. \cite{ladder},
the spin operators for the $n^{th}$ rung
and the two possible momenta in $y$-direction are:
\begin{eqnarray}
\bbox{S}(n,0) &=& 
\bbox{\tau}^\dagger_{n} \times \bbox{\tau}_n^{}
\label{spin0} \\
\bbox{S}(n,\pi) &=& \frac{1}{\sqrt{2}} (\bbox{\tau}^\dagger_n +
\bbox{\tau}_n^{} ).
\label{spinpi}
\end{eqnarray}  
The spin operator with transverse momentum $0$ thus actually corresponds
to a two-particle spectrum, and thus will take (in an infinite system)
the form
of a continuum, whereas the spin operator with transverse momentum
$\pi$ is a single particle-like spectrum. This difference
can indeed be seen in the numerical spin 
correlation function (Ref.\cite{ladder}).\\
The `$SO(5)$ partner' of the operator $\bbox{S}(n,\pi)$ therefore is
\begin{equation}
\tilde{\Delta}_n = \frac{1}{\sqrt{2}} (t_{n,1}^\dagger + t_{n,1}^{}),
\end{equation}
which is related to the creation and annihilation of a
singlet electron pair.
On the other hand, the operator of electron density with
transverse momentum $0$ is
the `$SO(5)$ partner' of $\bbox{S}(k,0)$: 
\begin{equation}
n_{n,0} = i ( t_{n,1}^\dagger  t_{n,5}^{} - 
t_{n,5}^\dagger  t_{n,1}^{}),
\label{den0}
\end{equation}
where $t_{n,1}^\dagger$ and $t_{n,5}^\dagger$ were defined in
(\ref{bosons}). 
Comparing (\ref{den0}), (\ref{spin0}) to (\ref{root}) we note that
the density and spin operators for transverse momentum transfer
$0$ are actually root generators of $SO(5)$ - it follows that we can
invoke the Wigner-Eckart-theorem (or intuitive arguments)
to show that the respective spin and density spectra for
certain pairs of states are identical. Below we will
present various examples.\\
As for the density operator
with transverse momentum transfer $\pi$, this is much more complicated
because this operator annihilates
any state corresponding to either charged or spinful Bosons.
This correlation function therefore will be determined by
the dynamics of the singly and threefold occupied rung, which
in our strong coupling limit is admixed only virtually.
We therefore expect that this correlation function has only weak
features at relatively high energies.\\
After these preliminaries we proceed to a discussion of the numerical
spectra.
With regards to the SCF it is important to keep in mind that all following
spectra have been computed using the 
$z$-component of the spin operator.\\
The half-filled ground state does not contain any Bosonic
excitation at all. We can obtain a nonvanishing
SCF only by coupling to the weakly admixed singly
and threefold occupied rung, and therefore expect
an extremely weak signal for $k_T$$=$$0$.
Final states for the SCF with $k_T$$=$$\pi$
are in the $(1,0)$ representation, and the dominant peak in the
respective spectra therefore gives the dispersion of the triplet Boson
(see Figure \ref{fig9}). 
Choosing now the lowest of these $(1,1)$ states,
the $^3(\pi,\pi)_0$, as the initial state for the
calculation of the SCF with $k_T$$=$$0$, the triplet Boson with
momentum $k$$=$$\pi$ could in principle be scattered into a state
with momentum $k+\pi$, where $k$ is the momentum transfer
along the ladder. Since we are using the $S_z$ operator, however,
this is not possible (see the representation (\ref{spin0}))
if we use the $S_z$$=$$0$ member of the $^3(\pi,\pi)_0$ state.
We therefore expect a very low intensity spectrum which again
is due to virtually admixed singly and three-fold occupied rungs.
Moreover, since the $z$-like triplet acts precisely like a hole
pair, and since the hole 
%%%%%%%%%%%%%%%%%%%%%%%%%%%%%%
\noindent
%%%%%%%%%%%%%%%%%%%%%%%%%%%%%%
\begin{figure}
\vspace{-0.5cm}
\epsfxsize=7cm
\epsffile{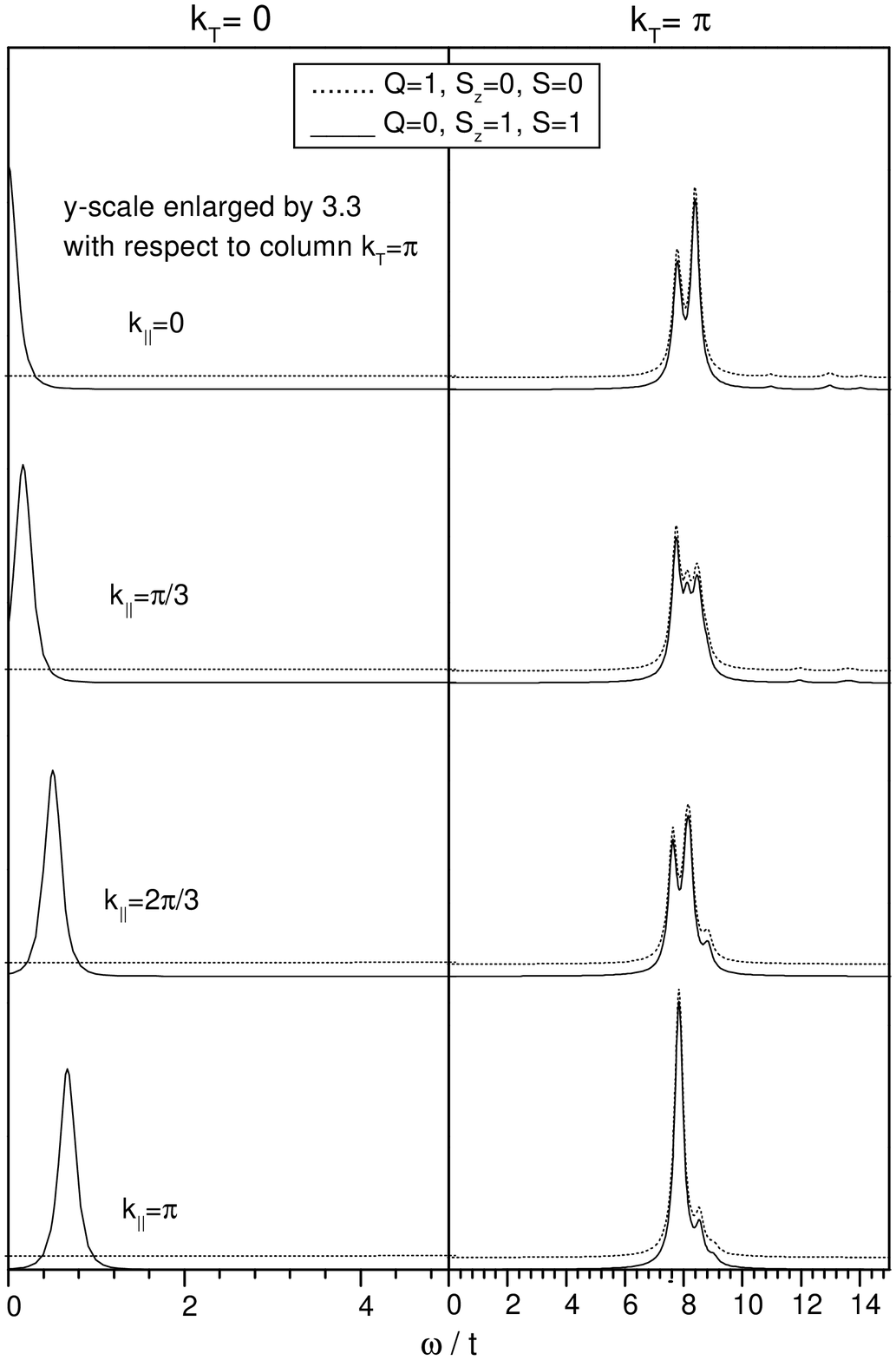}
\narrowtext
\caption{Comparison of the spin correlation spectrum for
the two-hole ground state (dotted line) and the half-filled
$^3(\pi,\pi)_0$ state with $S_z$$=$$1$ (full line).
The spectra for the two-hole state have been offset in
y-direction to facilitate the comparison. 
Parameter values are as in Figure 6.}
%SCZ%\ref{fig7}
\label{fig10}
\end{figure}
\noindent
%%%%%%%%%%%%%%%%%%%%%%%%%%%%%%%%%%%%%%%
pair cannot be excited by
the spin operator either, the $k_T$$=$$0$ spectra 
of the  $^3(\pi,\pi)_0$ state with $S_z$$=$$0$ and the
$^1(0,0)_{-1}$ ground state must be identical by $SO(5)$ symmetry
(which indeed they are, compare Figure \ref{fig11}).
The situation changes completely if we use the
$S_z$$=$$1$ member of the $^3(\pi,\pi)_0$ triplet as
initial state (see Figure \ref{fig10}): 
this state contains a mixture of the $x$-like and
$y$-like triplet, which both can be
excited by the $S_z$-operator with $k_T$$=$$0$.
We will therefore see a strong peak, with essentially the same dispersion
as the dominant peak in the $k_T$$=$$\pi$ spectra in Figure
\ref{fig9}, but shifted by $\Delta k$$=$$ \pi$ 
in momentum and by $\approx E_1-E_0$ in energy.
On the other hand acting with the $z$-like spin operator
with $k_T$$=$$\pi$ creates or annihilates a $z$-like
Boson. Then, in both states, $^3(\pi,\pi)_0$
with $S_z$$=$$1$ and $^1(0,0)_{-1}$, annihilation of
a $z$-like Boson is not possible. Creating
a $z$-like Boson, we will generate $SO(5)$ equivalent states, because
the created $z$-like Boson interacts with the one already
present in the system ($x$-like, $y$-like or hole-like) in identical ways
- the $k_T$$=$$\pi$ spectra for these states thus must be identical, 
which indeed they are (compare Figure \ref{fig10}).
Comparing Figure \ref{fig10} with Figure \ref{fig9} we see that the
magnon mode with $(\pi,\pi)$ at half-filling evolves continuously
into the $\pi$ resonance mode of the two hole state, in accordence 
with the analysis of reference \cite{S-Z-H}. Away from $(\pi,\pi)$,
a lower energy structure emerges in the SCF of the two hole state.\\
On the other hand, when we switch
to $^3(\pi,\pi)_0$ with $S_z$$=$$0$, 
the parts corresponding to a creation of the $z$-like Boson are 
slightly different from the spectra
in Figure \ref{fig10} because the final state
now contains two Bosons of equal ($z$-like) species, which interact
differently as compared to unequal species.\\
We proceed to a discussion of the DCF. In the remainder
of this section we will refer only to spectra with $k_T$$=$$0$.
It follows by $SO(5)$ symmetry and (\ref{spin0}), (\ref{den0}) that 
the density operator with $k_T$$=$$0$ takes a completely analogous form as
the spin operator. When using an $SO(5)$ singlet
as initial state, both operators thus must give the same
spectrum. This can be seen by comparing the SCF and DCF
for the half-filled ground state, compare Figure \ref{fig9} and
\ref{fig12}. They are also identical when acting onto states
which contain only $z$-like Bosons - which can be excited neither by the
$S_z$ or the density operator. 
This explains the identity
of the SCF and DCF for the $^3(\pi,\pi)$ state with $S_z$$=$$0$,
compare Figure \ref{fig9} and Figure \ref{fig12}.\\
On the other hand, since the hole pair behaves like a triplet,
the DCF for the two-hole ground state must be identical
to the SCF for the $^3(\pi,\pi)_0$ state with $S_z$$=$$1$,
compare Figures \ref{fig10} and \ref{fig12}, right panel.\\
Summarizing this section, we have seen that $SO(5)$ symmetry
enforces identity relationships between
spin and charge correlation functions for different
initial states, which differ by their hole number. Again this implies 
strong continuity with hole doping. 
While such continuity is certain to be present
in any system with (approximate) $SO(5)$ symmetry, it should be noted that 
the 
%%%%%%%%%%%%%%%%%%%%%%%%%%%%%%%%%%%%%%%%%
\begin{figure}
\vspace{-0.5cm}
\epsfxsize=7cm
\epsffile{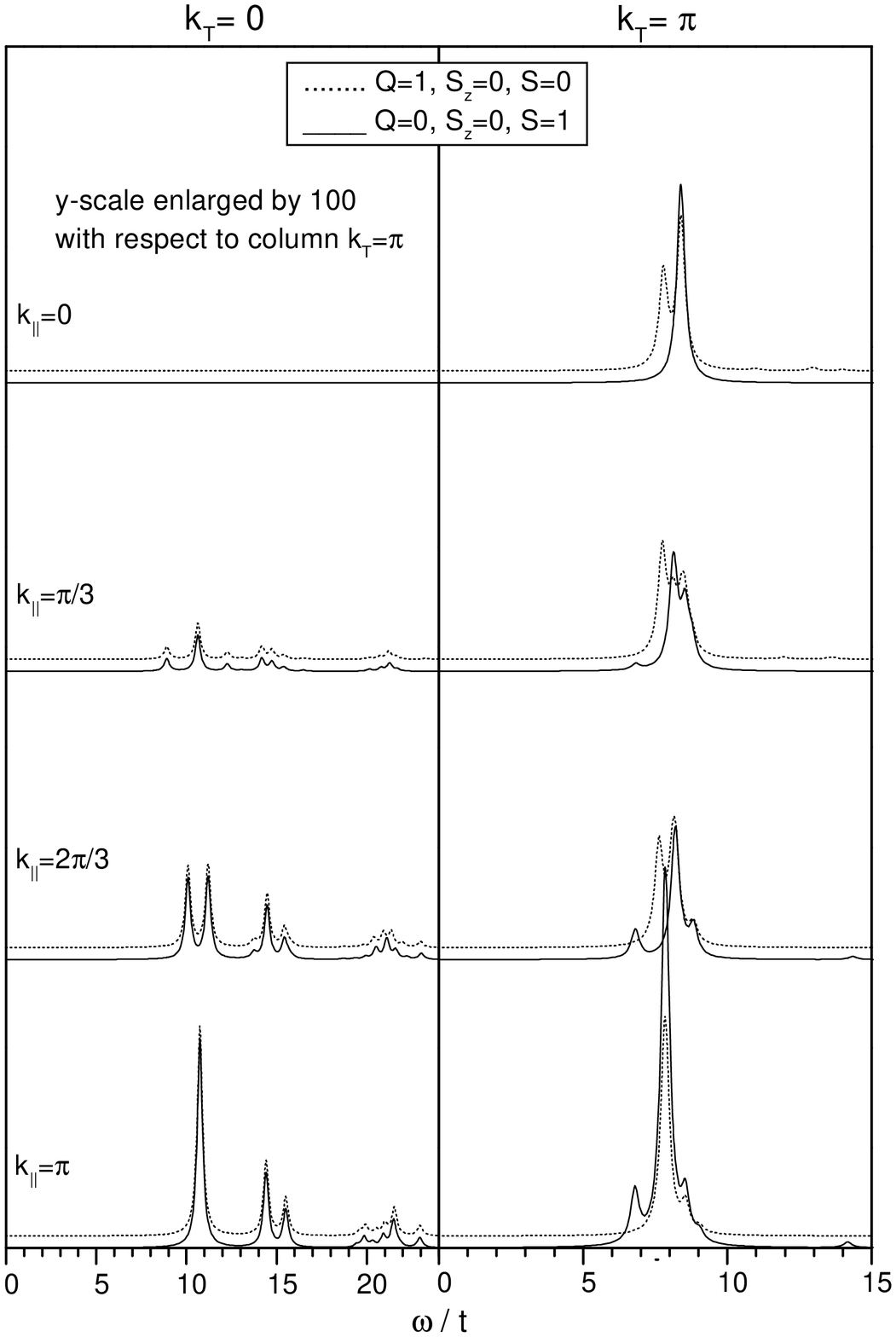}
\narrowtext
\caption{Comparison of the spin correlation spectrum for
the two-hole ground state (dotted line) and the
$^3(\pi,\pi)_0$ state with $S_z$$=$$0$ (full line). 
The spectra for the two-hole state have been offset in
y-direction to facilitate the comparison. 
Parameter values are as in Figure 6.}
%SCZ% \ref{fig6}
\label{fig11}
\end{figure}
%%%%%%%%%%%%%%%%%%%%%%%%%%%%%%%%%%%%%%%%%
\begin{figure}
\vspace{-1.5cm}
\epsfxsize=7cm
\epsffile{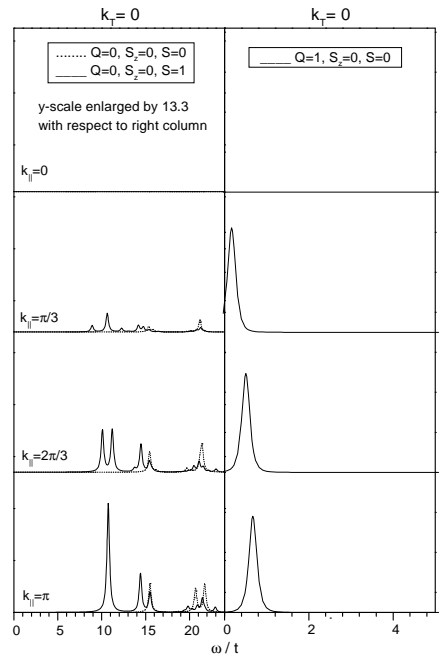}
\narrowtext
\caption{Comparison of the density correlation spectrum for
the half-filled ground state (dotted line, left), the
$^3(\pi,\pi)_0$ state with $S_z$$=$$0$ (full line, left), and
the two-hole ground state (right). Parameter values are as in
Figure 6.} 
\label{fig12}
\end{figure}
\noindent
%%%%%%%%%%%%%%%%%%%%%%%%%%%%%%%%%%%%%%%%%
identity relationships in the present case depend very much on
the representations (\ref{spin0}) and (\ref{den0})
of the spin and density operators in terms of the rung Bosons.
While these are easily transferable to ladder-like systems,
analogous identities for a fully planar system may be quite
different and more restricted.
%%%%%%%%%%%%%%%%%%%%%%%%%%%%%%%%%%%%%%%%%%%%%%
%SCZ%\section{Adiabatic connection to the {\small t}-J model}
\section{Adiabatic connection to the t-J model}
\begin{figure}
\vspace{-0.5cm}
\epsfxsize=7cm
\epsffile{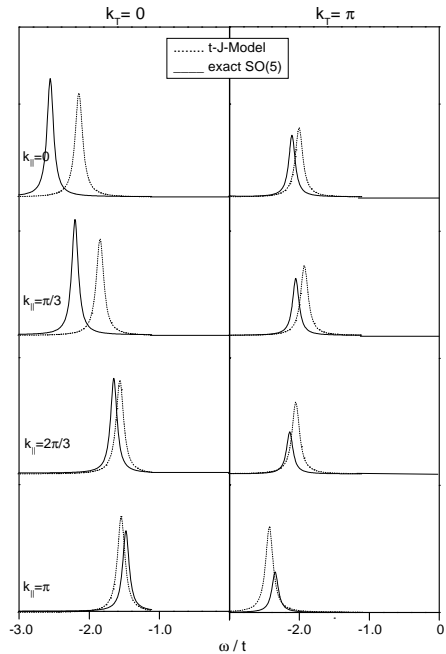}
\vspace{-0.25cm}
\narrowtext
\caption{Comparison of the photoemission spectra from 
the half-filled ground state of the $t-J$ model (dotted line)
and the exact $SO(5)$ model (full line). The $t-J$ parameters
are $t_\perp/t$$=$$0.5$, $J/t$$=$$0.5$, $J_\perp/t$$=$$2$ ,
the $SO(5)$ parameters $t_\perp/t$$=$$2/3$,
$U/t$$=$$4$, $V/t$$=$$3.25$, $J/t$$=$$3$, with $t_{SO(5)}$$=$$0.625 t_{t-J}$.}
\label{fig13}
\end{figure}

\begin{figure}
\vspace{-1.0cm}
\epsfxsize=7cm
\epsffile{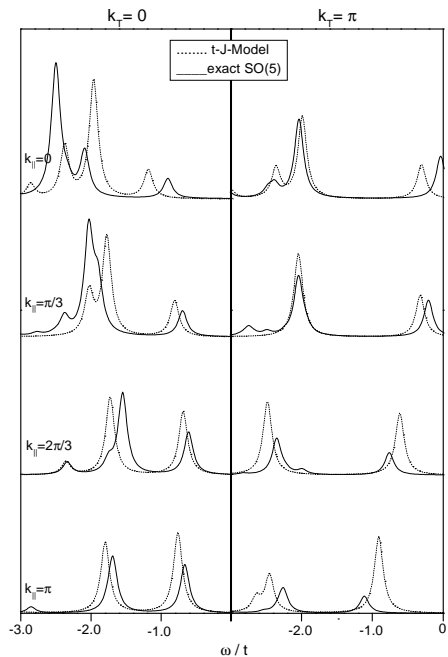}
\vspace{-0.25cm}
\narrowtext
\caption{Comparison of the photoemission spectra from 
the half-filled $^3(\pi,\pi)$ state with $S_z$$=$$0$
of the $t-J$ model (dotted line) and the exact $SO(5)$ model (full line). 
Parameters are as in Figure 13.}
%SCZ% \ref{fig16}
\label{fig14}
\end{figure}

\begin{figure}
\vspace{-0.5cm}
\epsfxsize=7cm
\epsffile{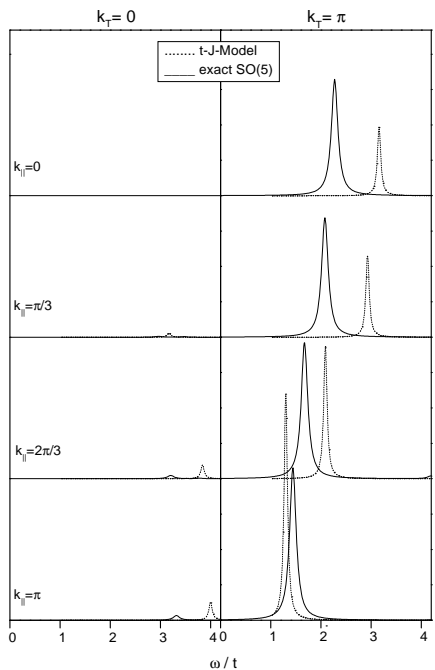}
\narrowtext
\caption{Comparison of the spin correlation spectra for 
the half-filled ground state
of the $t-J$ model (dotted line) and the exact $SO(5)$ model (full line). 
Parameters are as in Figure \ref{fig13}.}
\label{fig15}
\end{figure}

%%%%%%%%%%%%%%%%%%%%%%%%%%%%%%%%%%%%%%%%%%%%%%
So far, we have studied the spectra for the ideal $SO(5)$ symmetric
model. In this section we want to investigate the effect of
discarding the  correction terms $H_1$ which enforced exact $SO(5)$ symmetry.
More precisely our goal is to find, in the spirit of Landau's
Fermi liquid theory,
a `mapping' between the excitation spectrum of the exactly $SO(5)$
symmetric but in principle unphysical model, and that of a 
`physical' $t-J$ ladder. The latter, being the $U$$ \rightarrow$$\infty$
limit of the Hubbard model, 
incorporates the constraint of no double occupancy 
which, based on erroneous mean-field calculations,
has recently been argued to break $SO(5)$ symmetry\cite{Greiter}.
We compare the single particle spectra as well
as the spin correlation (i.e. a two particle spectrum)
of the $SO(5)$ symmetric ladder
and the $t-J$ model for a $6$-rung ladder. Thereby we adjust the parameters of
the $SO(5)$ symmetric ladder so as to obtain an optimal fit
to the $t-J$ model, for which we fix $t_\perp/t$$=$$0.5$ and $J/t$$=$$0.5$,
$J_\perp/t$$=$$2.0$. We are using a relatively large
value of the exchange along the rungs, $J_\perp$ - this is by no means
adequate to describe actual materials, but our main goal
here is to investigate the effect of the constraint
of no double occupancies in the $t-J$ model. Note that the
$t-J$ model has a `Hubbard gap' which is effectively infinite, whereas in the
$SO(5)$ symmetric ladder hole-doped
to electron-doped states are degenerate. The models
thus appear at first sight entirely unrelated, but our goal is 
to check our above conjecture that as long as we restrict ourselves to the
hole doped sector, the two models still can well have essentially
the same low-energy dynamics.\\
Figure \ref{fig13} and 
Figure \ref{fig14} show the photoemission spectra
originating from the half-filled 
ground state $^1(0,0)_0$, and $^3(\pi,\pi)_0$ with $S_z$$=$$0$, 
The overall agreement of the two model's spectra was obtained 
choosing $U/t$$=$$4$, $V/t$$=$$-3.25$, $J/t$$=$$3$, 
and $t_\perp/t$$=$$2/3$ for the exactly
$SO(5)$ symmetric model, followed by an energy rescaling by a factor of $0.625$
(which amounts to setting $t_{SO(5)} $$=$$ 0.625 t_{t-J}$).
The resulting mapping is excellent: the $t-J$ model's number of peaks, 
their weight distribution and dispersion are accurately obtained also
in the exactly $SO(5)$ symmetric model. In particular, the appearance of the
`sidebands' in the spectra for spin polarized ground states
can be nicely seen for both models.
Regarding the spin correlation for the same set of parameters 
(Figure \ref{fig15}), there is still good agreement as to the 
%%%%%%%%%%%%%%%%%%%%%%%%%%%%%%%%%%%%%
%%%%%%%%%%%%%%%%%%%%%%%%%%%%%%%%%%%%%
\noindent 
%%%%%%%%%%%%%%%%%%%%%%%%%%%%%%%%%%%%%
number and the weight of the peaks and the character 
of the dispersion.
The absolute size of the dispersion
is somewhat smaller in the exact $SO(5)$ model but the value of the
`spin gap' at $(\pi,\pi)$ is reproduced quite accurately.
Alltogether, the mapping between the `unphysical' 
but exactly $SO(5)$ symmetric model and the
physically better founded $t-J$ model is reasonably good, which clearly
supports the physical relevance of $SO(5)$ symmetry. 
We also note that the calculations are
`as close as possible' to half-filling, where the
impact of the Hilbert-space projection still is presumably the largest
in the $t-J$ model - yet the agreement of the low-energy physics
is obviously quite good. One therefore may hope
that similar mappings can be carried out also
for more realistic parameter values of $t-J$ ladders, so that
for $t-J$ and Hubbard ladders
the $SO(5)$ symmetric model may in fact be the generic effective
Hamiltonian. This is further supported by numerical
studies in the case of large $U$$=$$-V$,
$J$$=$$0$\cite{Duffy} recently carried out by Duffy, Haas and Kim, 
and renormalizatiion group 
calculations\cite{Enrico,Fisher} for the weak coupling case.
%%%%%%%%%%%%%%%%%%%%%%%%%%%%%%%%%%%%%%%%%%%%%%%%
\section{Conclusion}
%%%%%%%%%%%%%%%%%%%%%%%%%%%%%%%%%%%%%%%%%%%%%%%%
In summary, we have studied the dynamics of an exactly $SO(5)$ symmetric
ladder model. We have shown that in the strong-coupling limit the
model reduces to the simplest possible
$SO(5)$ symmetric generalization of a Bosonic Model
derived previously by
Gopalan {\em et al.}\cite{Gopalan} for the spin excitations of 
Heisenberg ladders. In this limit the
`vacuum' of the theory is the rung-RVB state, and
the elementary excitations of the $SO(5)$ symmetric ladder
correspond to uncharged triplet-like and charged singlet-like Bosons.
$SO(5)$ symmetry then simply states that the spin-like and
charge-like Bosons are dynamically equivalent, in the same
sense as e.g. proton and neutron are considered dynamically equivalent
in the isospin theory of nuclear physics.
In the strong coupling limit, the low energy physics can be mapped
onto a model of five species of hard-core bosons. The effective 
non-linear $\sigma$ model description\cite{science} can be 
systematically derived in this limit.\\
Because the ground state of the ladder system is a Mott insulator 
with `RVB type" of singlet vacuum, it can serve as a reference state
to consider the condensation of magnons and charged bosons on the equal
footing, thereby revealing the precise analogy between antiferromagnetism
and superconductivity. In a 2D system, no such reference state is found
to be the ground state 
for any reasonable Hamiltonian in the infinite system. However, the
ground state of any finite cluster is a total spin singlet, and recent
numerical calculations \cite{Meixner,ede97} demonstrate that the ground
state of the $t-J$ or Hubbard model is also approximately a $SO(5)$ 
singlet. The low energy excitations of the 2D cluster are magnons and
hole pairs, and share many similarities to the properties of the 
$SO(5)$ ladder system found in this work. \\
As shown in the present work, $SO(5)$ symmetry has profound
implications for the dynamical correlation functions, most
notably the single particle spectrum. Specific predictions
of $SO(5)$, like a `generalized rigid band 
behaviour'\cite{EderOhtaShimozato,Nishimoto}
and the appearance of sidebands in the inverse photoemission
spectrum\cite{inverse} may indeed have been observed long ago
in the actual 2D $t-J$ model. Motivated by the present theory we have
carried out more detailed spectroscopies on the 2D model
and obtained results in strong support of $SO(5)$\cite{tobepub}.\\
Finally we (and other authors\cite{Enrico,Fisher,Duffy})
have demonstrated that despite the at first sight rather 
unphysical parameter values of the $SO(5)$ symmetric model, a `Landau mapping' 
to the more realistic $t-J$ model is feasible. This may suggest that
the $SO(5)$ symmetric ladder is in fact the generic effective Hamiltonian
for two-leg ladder systems.\\
{\em Acknowledgements:}
Two authors (W.H. and S.C.Z.) have benefitted from many enlightening
discussions with D. J. Scalapino.
SCZ is supported in part by the NSF under grant numbers DMR94-00372
and DMR95-22915.
A.D., M.G.Z. and W.H. acknowledge support from DFN contract
TK598-VA/D3
and are grateful to the University of Stuttgart and
M\"unchen Supercomputing Centers.
R. E. most gratefully acknowledges the kind hospitality
at the National Center for Theoretical Studies,
National Tsing Hua University, Taiwan.
%%%%%%%%%%%%%%%%%%%%%%%%%%%%%%%%%%%%%%%%%%%%%%%%
\section{Appendix}
%%%%%%%%%%%%%%%%%%%%%%%%%%%%%%%%%%%%%%%%%%%%%%%%
We consider the coherent state
\[
|\Psi_\lambda \rangle = \frac{1}{\sqrt{n}}
e^{\lambda \sqrt{N} t_z(\pi)^\dagger}
 |vac\rangle,
\]
where $n$ is the normalization factor, and
$t_z(\pi)^\dagger$ the creation operator for a true
{\em hard-core} Boson.
Let us assume that the exponential has been expanded, and ask for the
total norm, $n_\nu$, of all states of $\nu^{th}$ order in $\lambda$
(i.e. all states containing $\nu$ Bosons).
Each of these terms has a prefactor of $\lambda^\nu/\nu!$ from the
expansion of the exponential. There is also
a factor of $\pm 1$, depending on how many Bosons are
on rungs with odd numbers - this will disappear upon squaring
the prefactor and we therefore disregard it. 
Moreover, the factor of $1/\nu!$ is cancelled
because each Boson configuration
is generated in $\nu!$ different ways when
expanding $(t_z(\pi)^\dagger)^\nu$. The number of different
Boson configurations (which are mutually orthogonal!)
with $\nu$ Bosons is $N!/(N-\nu)! \nu!$, whence 
\[
n_\nu = \lambda^{2\nu}  \left(
\begin{array}{c}
N\\
\nu\\
\end{array} \right),
\]
and summing over $\nu$ we obtain
\[
n = (1 + \lambda^2)^{N}.
\]
Next, we have
\[
\sqrt{N} t_z(\pi)^\dagger
| \Psi_\lambda \rangle = \partial_\lambda ( \sqrt{n} | \Psi_\lambda \rangle),
\]
whence
\begin{eqnarray*}
\langle \Psi_\lambda | M_S | \Psi_\lambda \rangle &=& \frac{1}{\sqrt{2}}
\partial_\lambda \log(n) 
\nonumber \\
&=& \frac{\sqrt{2} \lambda N}{1+ \lambda^2},
\end{eqnarray*}
q.e.d.
 
\end{multicols}
\end{document}